\documentclass{emulateapj}
\usepackage{placeins}
\usepackage{graphicx}
\usepackage{amsmath}
\usepackage{mathrsfs}
\shorttitle{Circumplanetary Disk Accretion}
\shortauthors{Zhu}

\begin{document}

\title{Accreting Circumplanetary Disks: Observational Signatures}

\author{Zhaohuan Zhu\altaffilmark{1,2}}

\altaffiltext{1}{Department of Astrophysical Sciences, 4 Ivy Lane, Peyton Hall,
Princeton University, Princeton, NJ 08544, USA}
\altaffiltext{2}{Hubble Fellow}

\email{zhzhu@astro.princeton.edu }

\newcommand\msun{\rm M_{\odot}}
\newcommand\lsun{\rm L_{\odot}}
\newcommand\msunyr{\rm M_{\odot}\,yr^{-1}}
\newcommand\be{\begin{equation}}
\newcommand\en{\end{equation}}
\newcommand\cm{\rm cm}
\newcommand\kms{\rm{\, km \, s^{-1}}}
\newcommand\K{\rm K}
\newcommand\etal{{\rm et al}.\ }
\newcommand\sd{\partial}
\newcommand\mdot{\rm \dot{M}}
\newcommand\rsun{\rm R_{\odot}}

\begin{abstract}
I calculate the spectral energy distributions (SEDs) of  accreting circumplanetary disks using
 atmospheric radiative transfer models. 
Circumplanetary disks only accreting at  $10^{-10}\msunyr$ around a 1 M$_{J}$ planet can be brighter than the planet itself.
 A moderately accreting circumplanetary disk ($\dot{M}\sim 10^{-8}\msunyr$; enough to form a 10 M$_{J}$ planet 
 within 1 Myr) around a 1 M$_{J}$ planet has a maximum temperature of $\sim$2000 K, 
 and at near-infrared wavelengths ($J$, $H$, $K$ bands),
this disk is as bright as a late M-type brown dwarf or a 10 M$_{J}$ planet with a ``hot start''. 
To use direct imaging to find  the accretion disks around low mass planets (e.g., 1 M$_{J}$) and distinguish them
from brown dwarfs or hot high mass planets, it is crucial  to obtain photometry at mid-infrared bands ($L'$, $M$, $N$ bands) 
because the emission from circumplanetary disks falls off more slowly towards longer wavelengths than those of brown dwarfs or planets.
If  young planets have strong magnetic fields ($\gtrsim$100 G), fields may truncate slowly accreting circumplanetary disks ($\dot{M}\lesssim$10$^{-9}\msunyr$) 
and lead to magnetospheric accretion,
which can provide additional accretion signatures, such as  UV/optical excess from the accretion shock and line emission.

\end{abstract}


\section{Introduction}
 {Planets form and grow in circumstellar disks.} Before a planet's mass reaches to
the mass of Jupiter, its tidal force opens a gap in the circumstellar disk 
(Lin \& Papaloizou 1986, Kley \& Nelson 2012 and references therein). Material
that resides beyond the gap in the circumstellar disk can be accreted by the protoplanet, but
 forms a circumplanetary disk because of its large angular momentum (Lubow \etal 1999; Ayliffe \& Bate 2009). 
The accretion of the circumplanetary disk determines the final mass of the giant planet. 
Since circumstellar disks have lifetimes of a few million years  (Hernandez \etal 2007), forming a 1$-$10 M$_{J}$ planet within such time
requires that the circumplanetary disk has an average accretion rate of $\dot{M}\gtrsim 10^{-9}-10^{-8} \msunyr$.

Although there have been no clear detections of circumplanetary disks yet (except for a possible candidate, \S 5.3),
theoretical studies show the possibility of complex circumplanetary disk structures: the infall from circumstellar to circumplanetary
disks can occur at high altitudes (Tanigawa \etal 2012, Szulagyi \etal 2014), the infall can carry little angular momentum (Canup \& Ward 2002),
the infall can be episodic (Gressel \etal 2013),
the surface of the disk can be subject to magnetorotational instability (MRI) (Fujii \etal 2011, 2014, Turner \etal 2014), a magnetocentrifugal wind can develop
in the disk (Gressel \etal 2013), the disk midplane can be dominated by Hall
MHD (Keith \& Wardle 2014), and the disk can undergo outbursts (Lubow \& Martin 2012).

Future circumplanetary disk observations are needed to 
constrain their structure. Fortunately, finding circumplanetary disks may not be too difficult. 
A disk around a 1 M$_{J}$ planet accreting at $\dot{M}=10^{-8}\msunyr$ has an
accretion luminosity of
\begin{equation}
L_{disk}=\frac{G{\rm M}_{J}\dot{M}}{2 {\rm R}_{J}}=1.5\times10^{-3} {\rm L}_{\odot}\,,
\end{equation}
which is as bright as a late M-type/early L-type brown dwarf (Basri 2000; Chabrier 
\& Baraffe 2000). 
Finding these disks in future observations will enable us to 
test accreting disk theory, constrain satellite formation, and finally find young planets. 

In this paper  theoretical SEDs of accreting circumplanetary disks are calculated. These SEDs
together with other accretion signatures
can be used for planning future observations to discover such disks.
In \S 2,  the radiative transfer model is introduced. The SEDs are shown in \S 3.
The observational signatures of magnetospheric accretion are presented in \S 4.
Finally, after a short discussion in \S 5, the paper is
summarized in \S 6.

\section{Method}
As a first attempt to calculate the SEDs of accreting circumplanetary disks, 
I only focus on disks whose accretion luminosity is stronger than the planet irradiation.
This requirement significantly simplifies the SED calculation since the SEDs of these disks are independent of the planet properties.
To be more specific, I only study
accreting circumplanetary disks with $\dot{M}\gtrsim 10 ^{-10}\msunyr$ around 
low mass planets, such as a 1 M$_{J}$ planet. A 1 M$_{J}$ planet has an effective temperature 
of $\sim$800 K at an age of 1 million years in the ``hot start'' planet model 
(Spiegel \& Burrows 2012, SB), and it has a total luminosity
of 4$\times 10^{-6}$ L$_{\odot}$. Meanwhile, the luminosity of an accretion disk with 
$\dot{M}\gtrsim 10 ^{-10}\msunyr$ is $\gtrsim$ $10^{-5}$L$_{\odot}$. Thus,
the irradiation from the planet to the disk can be ignored
in the disk SED calculation. For slowly accreting circumplanetary disks (e.g. $\dot{M}\sim 10 ^{-10}\msunyr$) 
around high mass planets with ``hot start'' (e.g. a 10 M$_{J}$ planet with an effective temperature of 2000 K), a proper treatment
including the planet irradiation is needed and left for future publications. 
This viscous heating dominated disk resembles FU Orionis systems for protostars (Hartmann \& Kenyon 1996).
Thus,
I follow the method of Zhu \etal (2007, 2008, 2009) and Calvet \etal (1991a,b) (which was used to calculate
the SEDs of FU Orionis systems) to 
calculate the disk
spectrum. In summary, I calculate the emission from the atmosphere of
 a viscous, geometrically thin,
optically thick accretion disk with constant mass accretion rate
$\dot{M}$ around a planet with mass $M_{p}$ and radius $R_{p}$. The disk height
$H$ is assumed to vary with the distance from the 
 planet as $H=H_{0}(R/R_{in})^{9/8}$, where $H_{0}=0.1
R_{in}$ is assumed, and $R_{in}$ is the disk inner radius. This
approximation is not very accurate but it only affects the local
surface gravity of the disk atmosphere, which has only a small
effect on the emergent spectrum. We assume that
 radiative equilibrium holds in the disk
atmosphere, and the surface flux is determined by the viscous energy
generation in the deeper disk layers.  This constant radiative flux
through the disk atmosphere can be characterized by the effective
temperature distribution of the steady optically-thick disk as in 
\begin{equation}
\sigma T_{eff}^4 = {3 G M_{p}\dot{M} \over 8\pi\sigma R^{3}}
\left(1-\left(\frac{R_{in}}{R}\right)^{1/2}\right)\,,
\label{eq:Fv}
\end{equation}
where $M_p$ is the central planet's mass.
This equation predicts that the maximum disk temperature $T_{max}$
occurs at $1.36 R_{in}$ and then decreases to zero at $R = R_{in}$.  
Since this decrease of
temperature towards the planet is sensitive to the boundary condition at the planet surface,
Equation (\ref{eq:Fv}) is modified so
that, when the radius is smaller than 1.36 $R_{in}$, 
the temperature is constant and equal to $T = T_{max}$. 
The vertical temperature structure at each
radius is calculated using the gray-atmosphere approximation in the
Eddington limit, adopting the Rosseland mean optical depth $\tau$.

The emission from either the boundary layer or the magnetospheric accretion shock
has been neglected in this calculation. Their  contributions to the SED will be included in \S 4. Generally, when the magnetosphere is equal or smaller than 2 R$_{J}$, 
such emission would emerge in UV/optical and irrelevant
to modeling the infrared SED here, except that the accretion shock might enhance the heating
to the disk via irradiation (\S 4).

The opacity of atomic and molecular lines has been calculated using
the Opacity Distribution Function (ODF) method \citep{kurucz04,
kurucz042,castelli05}. The ODF method is a statistical
approach to handle line blanketing when millions of lines are
present in a short wavelength range \citep{kurucz74}. 
The line list is taken from \cite{kurucz05}.  Not only atomic lines but also many molecular
lines are included. The opacities of TiO and H$_{2}$O, the most
important molecules in the infrared, are from \cite{PS97} and
\cite{S98}. The details
on this method and opacity sources can be found in Zhu \etal (2007).
We improve the dust opacity in Zhu \etal (2007) by following D'Alessio \etal (2001). We assume that MgFeSiO$_{4}$ (Olivine)
and Mg$_{0.8}$Fe$_{0.2}$SiO$_{3}$ (Pyroxene) have dust-to-gas mass ratio of 0.0017 respectively, and
graphite has mass ratio of 0.0041.  The grain size distribution has a power law of 3.5 with 0.005$\mu$m and 1$\mu$m
as the minimum and maximum size.
The dust sublimation temperatures  for different dust species are taken from 
D'Alessio \etal (2001). At low temperatures complex chemical processes occur which are not included in the Kurucz data.
Low temperature
molecular opacity has not been calculated in detail;
instead, the abundance ratio between different types
of molecules below 700 K is assumed to be the same as the ratio at 700 K. This is unimportant
for our purposes because dust opacity dominates at such low temperatures.

The total flux from the accretion disk is the addition of the fluxes coming from all the annuli in the disk.
The radii of these annuli are chosen to
increase exponentially from $R = R_{in}$ to $R_{out}$. Since the size of the planet is uncertain, 
depending on the ``cold'' and ``hot start'' planet models (Marley et al. 2007, 1-2 R$_{J}$ in SB), and the inner disk may be truncated by the planet's magnetosphere (\S 4),
$R_{in}$ is varied from 1 to 4 Jupiter radii in our calculation. The outer radius $R_{out}$ is also very uncertain. Theoretically,
 $R_{out}$ should be smaller than the maximum extent of the circumplanetary disk $\sim$0.4 $R_{H}$ where $R_{H}$ is the Hill
radius of the planet (Martin \& Lubow 2011). 
For a Jupiter mass
planet at 20 AU, 
 this is around $\sim$ $1000$ R$_J$. Another rough constraint for $R_{out}$ can be derived if we assume that
the accretion in circumplanetary disks  is due to MRI.
The disk needs to be sufficiently ionized to sustained MRI and thermal collision can sufficiently ionize the disk up to 
50 R$_J$ (Keith \& Wardle 2014) when the disk accretes at 10$^{-8}\msunyr$. 
Furthermore, if material from the circumstellar disk  falls directly to the inner regions of the 
circumplanetary disk (Machida \etal 2008, Machida 2009, Tanigawa \etal 2012), the accretion disk will also be small.
{ However, while accreting, the circumplanetary disk will also expand to conserve angular momentum. Eventually the tidal
torque from the central star can remove the angular momentum of the circumplanetary disk when the 
disk extends to 0.4 $R_{H}$ (Martin \& Lubow 2011). In this case, the disk temperature structure deviates
from a standard viscous model and will be discussed in \S 5.1. Nevertheless, in the next section, we set $R_{out}$ to be 50 $R_{in}$ and 
1000 $R_{in}$ respectively to
demonstrate how insensitive the disks' SEDs depend on $R_{out}$.}

\section{SEDs of Accreting Circumplanetary Disks}
\begin{figure*}[ht!]
\centering
\includegraphics[trim=0cm 0cm 0cm 0cm, width=0.9\textwidth]{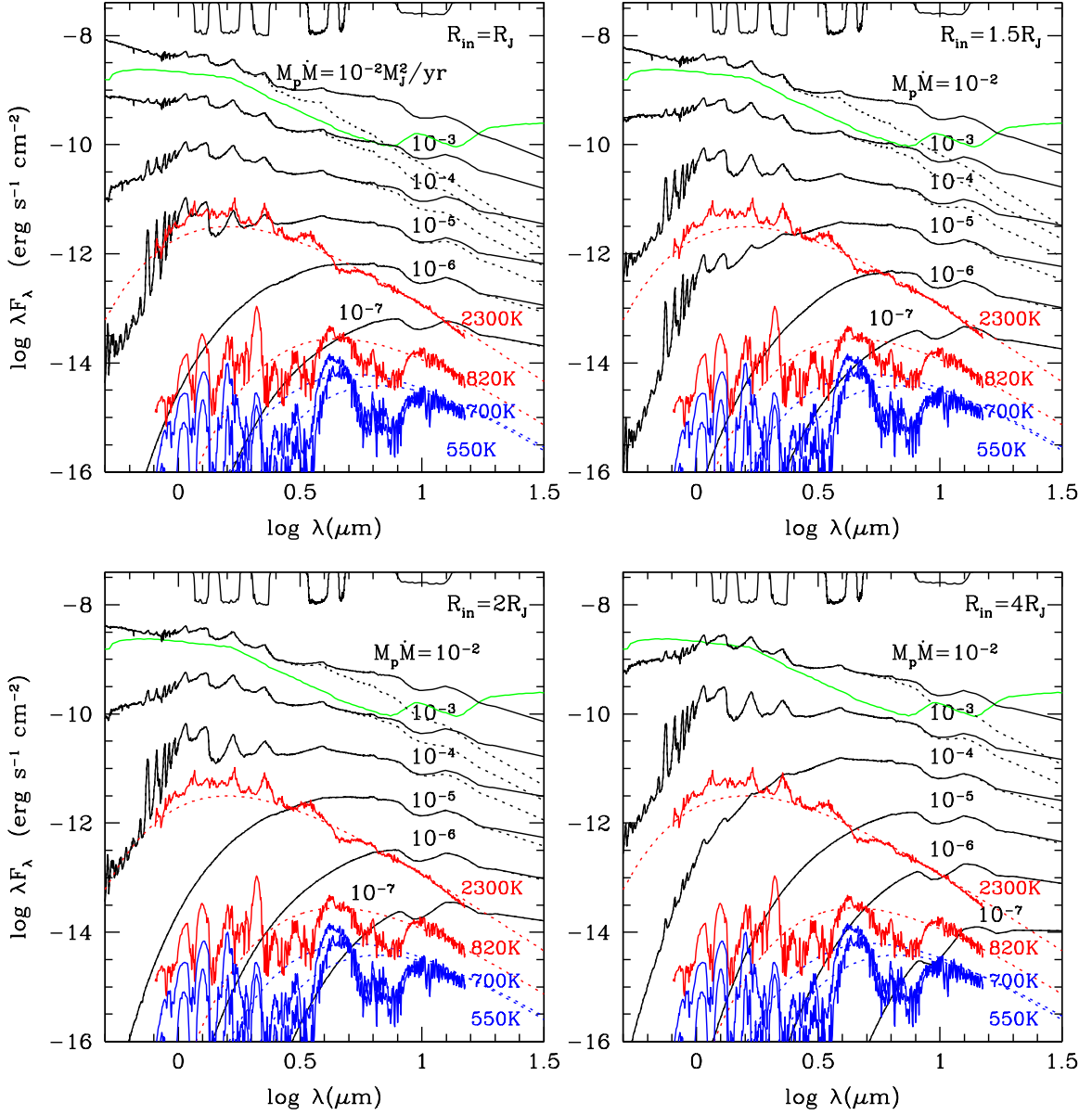} 
\caption{ The SEDs of accreting circumplanetary disks at 100 pc (black curves) with different disk inner radii ($R_{in}$). 
The solid curves represent the cases with $R_{out}=1000 R_{in}$ while the dotted curves are calculated with   $R_{out}=50 R_{in}$.
The product of the planet mass and the disk accretion rate ranges from 10$^{-7}$ to 10$^{-2}$ M$_{J}^{2}$yr$^{-1}$.
 For comparison, the red curves are the SEDs of the 1 Myr old planets at 100 pc based on 
the ``hot  start'' models (SB). The red curve with a brighter flux is from a 10 M$_{J}$ planet 
while the red curve with a weaker flux is from a 1 M$_{J}$ planet. We have also plotted the SEDs based on the ``cold start'' planet models
as the blue curves (SB). Similarly, the blue curve with a brighter flux is from a 10 M$_{J}$ planet 
while the blue curve with a weaker flux is from a 1 M$_{J}$ planet.
Since SB only gives the spectra from 0.8-15 $\mu$m, we also plot the SEDs from the blackbody having the corresponding planet size and
effective temperature (labeled along the curves) as the dotted color curves. For another comparison,
the green curve is the SED of the protostar GM Aur (model spectrum from Zhu \etal 2012) scaled to 100 pc. 
At the top of each panel, the black curves indicate the transmission
functions of $J$, $H$, $K$, $L'$, $M$, and $N$ bands. }
\vspace{-0.1 cm} \label{fig:SED}
\end{figure*}

The SED of a steady, optically-thick accretion disk is determined by
two parameters: the product of the mass of the planet and the disk accretion
rate  ($M_{p}\dot{M}$), and the disk inner radius ($R_{in}$).

Thus, we varied both $M_{p}\dot{M}$ and $R_{in}$ and the resulting 
SEDs of accreting circumplanetary disks are shown in Figure \ref{fig:SED} as the black curves, assuming that the disk is observed face on.
The solid curves represent the cases with $R_{out}=1000 R_{in}$ while the dotted curves are calculated with   $R_{out}=50 R_{in}$.
The distance to the object is assumed to be 100 pc which is the typical distance to the closest star forming region.
For comparison, the red curves are the SEDs of the 1 Myr old planets at 100 pc based on 
the ``hot  start'' models (SB). The red curve with a brighter flux is from a 10 M$_{J}$ planet 
while the red curve with a weaker flux is from a 1 M$_{J}$ planet. We have also plotted the SEDs based on the ``cold start'' models
as the blue curves (SB). Similarly, the blue curve with a brighter flux is from a 10 M$_{J}$ planet 
while the blue curve with a weaker flux is from a 1 M$_{J}$ planet. 
These planet models correspond to the 1 M$_{J}$,
10 M$_{J}$, ``hot/cold start'' models at  an age of 1 Myr in Table 1 and Figure 6 of SB. Since SB only gives the 
spectra from 0.8-15 $\mu$m, we also plot the SEDs from the blackbody having the corresponding planet size and
effective temperature (temperatures are labeled along the curves) as the colored dotted curves. For another comparison,
the green curve is the SED of the protostar GM Aur (model spectrum from Zhu \etal 2012) scaled to 100 pc.

The peak
of the SED (after reddening correction) and/or  spectral lines are mainly determined by 
the maximum temperature of the steady disk model
\begin{eqnarray}
T_{max}&=&0.488 \left ( \frac{3GM_{p}\dot{M}}{8\pi R_{in}^{3}\sigma} \right
)^{1/4}\nonumber\\
&\sim& 2257\, {\rm K} \left(\frac{M_{p}}{{\rm M}_{J}}\right)^{1/4}\left(\frac{\dot{M}}{10^{-8}\msunyr}\right)^{1/4}\left(\frac{R_{in}}{{\rm R}_{J}}\right)^{-3/4}
\,.\label{eq:tmax}
\end{eqnarray}
Thus, a disk around a 1 M$_{J}$ planet accreting at $10^{-8}\msunyr$  (labeled with $M_{p}\dot{M}=10^{-5}$M$_{J}^{2}$/yr in the upper left panel of Figure \ref{fig:SED})
has a similar optical and near-IR SED as the 2300 K blackbody (e.g., a late M-type brown
 dwarf, Burrows \etal 2001, or a 10 M$_{J}$ planet with a ``hot start'' , SB). 

The true luminosity
of a flat disk $L_{disk}$ is determined by
\begin{eqnarray}
L_{disk}&=& 2 \pi d^{2} {F \over \cos i} = { G M_{p} \dot{M} \over 2
R_{in}}\label{eq:ldisk}\\
&\sim& 1.46\times 10^{-3} \,{\rm L_{\odot}} \left(\frac{M_{p}}{{\rm M}_{J}}\right)\left(\frac{\dot{M}}{10^{-8}\msunyr}\right) \left(\frac{R_{in}}{{\rm R}_{J}}\right)^{-1}
\,,\label{eq:Ld}
\end{eqnarray}
%correct the equation
where $d$ is the distance to the system,
$i$ is the inclination angle of the disk to the line of sight, and
$F$ is the observed total flux corrected for extinction. Thus, accreting circumplanetary
disks can be quite bright. A disk accreting at $10^{-8}\msunyr$ is as bright as a late-M/early-L type brown dwarf.

 Figure \ref{fig:SED} shows that, at near to mid IR, the accreting circumplanetary disk is normally brighter than a 1 Myr old 1 M$_{J}$ planet no matter whether
 the planet is ``cold'' or ``hot start''
  as long as  $M_{p}\dot{M}\gtrsim 10^{-7}$M$_{J}^{2}$ yr$^{-1}$. Furthermore, the disk spectrum is redder than a single 
 blackbody spectrum, providing a way to distinguish the accretion disk from
 a planet or a background star. It also demonstrates that direct imaging at 
 longer wavelengths will provide a higher contrast ratio for the circumplanetary disk with respect to the planet.

\begin{table*}
\begin{center}
\caption{Absolute Multi-Band Magnitudes of Accreting Circumplanetary Disks \label{tab1}}
\begin{tabular}{cc|c|cccccc|cccccc}
\tableline\tableline
 &&&\multicolumn{6}{c|}{Full Disk}&\multicolumn{6}{|c}{Truncated Disk}\\
$M\dot{M}(M_{J}^2/yr)$ &  $R_{in}(R_{J})$ &  $T_{max}$ (K)&  J  & H & K & L' & M & N & J & H & K & L' & M & N \\ 
\tableline
10$^{-2}$ & 1  & 12542 & 2.3 & 2.0 & 1.7 & 0.7 & 0.2 & -1.2 & 2.3 & 2.0 & 1.7 & 1.3 & 1.4 & 1.1 \\
10$^{-3}$  & 1 & 7053 & 4.3 & 3.9 & 3.5 & 2.6 & 2.2 & 0.7 & 4.3 & 3.9 & 3.5 & 2.7 & 2.4 & 1.9 \\
10$^{-4}$  & 1 & 3966 & 6.5 & 6.0 & 5.6 & 4.5 & 4.1 & 2.6 &6.5 & 6.0 & 5.6 & 4.5 & 4.1 & 3.0 \\
6$\times$10$^{-5}$  & 1 & 3491 & 7.0 & 6.6 & 6.1 &4.9 & 4.5 & 3.0 &7.0 & 6.6 & 6.1 & 4.9 & 4.5 & 3.3 \\
3$\times$10$^{-5}$  & 1 & 2935 & 7.8 & 7.5 & 6.9 & 5.6 & 5.1 & 3.6 & 7.8 & 7.5 & 6.9 & 5.6 & 5.1 & 3.8\\
10$^{-5}$  & 1 & 2230 & 9.3 & 9.2 & 8.3 & 6.6 & 6.0 & 4.5 & 9.3 & 9.2 & 8.3 & 6.6 & 6.0 &  4.5\\
6$\times$10$^{-6}$  & 1 & 1963 & 10.1 & 9.9 & 8.9 & 7.1 & 6.4 & 4.9 & 10.1 & 9.9 & 8.9 & 7.1 & 6.4 & 4.9\\
3$\times$10$^{-6}$  & 1 & 1651 & 13.4 & 11.6 & 10.0 & 7.7 & 7.0 & 5.4 & 13.4 & 11.6 & 10.0 & 7.7 & 7.0 & 5.5 \\
10$^{-6}$  & 1 & 1254 & 16.0 &13.6 & 11.6 & 8.9 & 8.0 & 6.3 & 16.0 & 13.6 & 11.6 & 8.9 &8.0 & 6.3 \\
6$\times$10$^{-7}$  & 1  & 1104 & 17.7  & 15.0 & 12.6 & 9.5 & 8.6 & 6.8 & 17.7  & 15.0 & 12.6 & 9.5 & 8.6 & 6.8 \\
3$\times$10$^{-7}$  & 1 & 928 & 20.6 & 17.3 & 14.4 & 10.6 & 9.5 & 7.4 & 20.6 & 17.3 & 14.4 & 10.6 & 9.5 & 7.4\\
10$^{-7}$ & 1 & 705 &  25.2 & 20.8& 17.2 & 12.7 & 11.4 & 8.4 & 25.2 & 20.8 &17.2 & 12.7 & 11.4 & 8.4 \\ 
\tableline
10$^{-2}$ & 1.5  & 9253 & 2.5 & 2.0 & 1.7 & 0.8 & 0.4 & -1.1 & 2.5 & 2.0 & 1.8 & 1.1 & 1.0 & 0.6 \\
10$^{-3}$  & 1.5 & 5204 & 4.5 & 4.0 & 3.6 & 2.7 & 2.3 & 0.8 & 4.5 & 4.0 & 3.6 & 2.7 & 2.4 & 1.6 \\
10$^{-4}$  & 1.5 & 2926 & 6.9 & 6.5 & 6.0 & 4.7 & 4.2 & 2.7 & 6.9 & 6.5 & 6.0 & 4.7 & 4.2 & 2.9 \\
6$\times$10$^{-5}$  & 1.5 & 2575 &  7.5 & 7.3 & 6.6 & 5.1 & 4.6 & 3.1 & 7.5 & 7.3 & 6.6 & 5.1 & 4.6 & 3.2\\
3$\times$10$^{-5}$  & 1.5 & 2166 & 8.5 & 8.4 & 7.5 & 5.8 & 5.2 & 3.7 & 8.5 & 8.4 & 7.5 & 5.8 & 5.2 & 3.7 \\
10$^{-5}$  & 1.5 & 1646 & 11.7 & 10.6 & 9.1 & 6.9 & 6.2 & 4.6 & 11.7 & 10.6 & 9.1 & 6.9 & 6.2 & 4.6 \\
6$\times$10$^{-6}$  & 1.5  & 1448 & 13.6 & 11.6 & 9.8 & 7.4 & 6.6 & 5.0 & 13.6 & 11.6 & 9.8  & 7.4 & 6.6 & 5.0\\
3$\times$10$^{-6}$  & 1.5 & 1218 & 15.4 &13.0 & 10.9 & 8.1 & 7.3 & 5.6 & 15.4 &13.0 & 10.9 & 8.1 & 7.3 & 5.6 \\
10$^{-6}$  & 1.5 & 925 & 19.8 &16.4 & 13.6 & 9.8 & 8.6 & 6.5  & 19.8 &16.4 & 13.6 & 9.8 & 8.6 & 6.5 \\
6$\times$10$^{-7}$  & 1.5 & 814 & 22.0 & 18.2 & 15.0 & 10.9 &9.5 & 7.0 & 22.1 & 18.2 & 15.0 & 10.9 &9.5 & 7.0 \\
3$\times$10$^{-7}$  & 1.5 & 685 &  24.8 & 20.3 & 16.6 & 12.1 & 10.7 & 7.7 & 24.8 & 20.3 & 16.6 & 12.1 & 10.7 & 7.7 \\
10$^{-7}$ & 1.5 & 520 & 31.2 & 25.2 & 20.4 & 14.3 & 12.5 & 8.8 & 31.2 & 25.2 & 20.4 & 14.3 & 12.5 & 8.8 \\ 
\tableline
10$^{-2}$ & 2  & 7458 & 2.6 & 2.1 & 1.8 & 0.9 & 0.5 & -1.1 & 2.6 & 2.1 & 1.8 & 1.0 & 0.8 & 0.3 \\
10$^{-3}$  & 2 & 4194 & 4.7 & 4.2 & 3.8 & 2.8 & 2.4 & 0.9 & 4.7 & 4.2 & 3.8 & 2.8 & 2.4 & 1.4\\
10$^{-4}$  & 2 & 2358 & 7.3 & 7.1 & 6.4 & 4.8 & 4.3 & 2.8 & 7.3 & 7.1 & 6.4 & 4.8 & 4.3 & 2.9 \\
6$\times$10$^{-5}$  & 2 & 2076 & 8.1 & 7.9 & 7.0 & 5.3 & 4.7 & 3.2 & 8.1 & 7.9 & 7.0 & 5.3 & 4.7 & 3.2 \\
3$\times$10$^{-5}$  & 2 & 1745 & 9.5 & 9.1 & 8.0 & 6.0 & 5.3 & 3.8 &  9.5 & 9.1 & 8.0 & 6.0 & 5.3 & 3.8\\
10$^{-5}$  & 2 & 1326 & 13.8 & 11.6 & 9.7 & 7.2 & 6.4 & 4.7 & 13.8 & 11.6 & 9.7 & 7.2 & 6.4 & 4.7  \\
6$\times$10$^{-6}$  & 2 & 1167 & 15.4 & 12.8 & 10.7 & 7.7 & 6.9 & 5.1 & 15.4 & 12.8 & 10.7 & 7.7 & 6.9 & 5.1\\
3$\times$10$^{-6}$  & 2 & 981 & 18.1 & 15.0 & 12.3 & 8.7 & 7.7 & 5.7 &  18.1 & 15.0 & 12.3 & 8.7 & 7.7 & 5.7\\
10$^{-6}$  & 2 & 746 & 22.8 &18.6 & 15.2 & 10.9 & 9.5 & 6.7 &  22.8 &18.6 & 15.2 & 10.9 & 9.5 & 6.7 \\
6$\times$10$^{-7}$  & 2 & 656 &  25.0 & 20.3 & 16.5 & 11.7 & 10.3 & 7.2 & 25.0 & 20.3 & 16.5 & 11.7 & 10.3 & 7.2 \\
3$\times$10$^{-7}$  & 2 & 552 & 29.1 & 23.4 & 18.8 & 13.1 & 11.4 & 7.9 & 29.1 & 23.4 & 18.8 & 13.1 & 11.4 & 7.9 \\
10$^{-7}$ & 2 & 419 & 37.1 & 29.5 & 23.5 & 15.9 & 13.7 & 9.2 & 37.1 & 29.5 & 23.5 & 15.9 & 13.7 & 9.2 \\ 
\tableline
10$^{-2}$ & 4  & 4434 & 2.9 & 2.4 & 2.0 & 1.1 & 0.7 & -0.8 & 2.9 & 2.4 & 2.0 & 1.1 & 0.7 & -0.3 \\
10$^{-3}$  & 4 & 2494 & 5.4 & 5.2 & 4.5 & 3.1 & 2.6 & 1.1 & 5.4 & 5.2 & 4.5 & 3.1 & 2.6 & 1.2 \\
10$^{-4}$  & 4 & 1402 & 11.5 & 9.6 & 7.8 & 5.3 & 4.6 & 3.0 & 11.5 & 9.6 & 7.8 & 5.3 & 4.6 & 3.0 \\
6$\times$10$^{-5}$  & 4 &  1234 & 13.2 & 10.8 & 8.7 & 6.0 & 5.2 & 3.4 & 13.2 & 10.8 & 8.7 & 6.0 & 5.2 & 3.4  \\
3$\times$10$^{-5}$  & 4 & 1038 & 15.6 & 12.7 & 10.2 & 6.9 & 5.9 & 4.0 & 15.6 & 12.7 & 10.2 & 6.9 & 5.9 & 4.0 \\
10$^{-5}$  & 4 & 789 & 20.4 & 16.4 & 13.2 & 9.1 & 7.7 & 5.0 & 20.4 & 16.4 & 13.2 & 9.1 & 7.7 & 5.0  \\
6$\times$10$^{-6}$  & 4 & 694 &  22.4 & 18.0 & 14.4 & 9.8 & 8.5 & 5.5 & 22.4 & 18.0 & 14.4 & 9.8 & 8.5 & 5.5\\
3$\times$10$^{-6}$  & 4 & 584 &  26.2 & 20.8 & 16.5 & 11.1 & 9.5 & 6.2 & 26.2 & 20.8 & 16.5 & 11.1 & 9.5 & 6.2\\
10$^{-6}$  & 4 & 443 & 33.8 & 26.6 & 20.9 & 13.8 & 11.7 & 7.4 &  33.8 & 26.6 & 20.9 & 13.8 & 11.7 & 7.4 \\
6$\times$10$^{-7}$  & 4 & 390 & 38.1 & 29.9 & 23.4 & 15.3 & 12.9 & 8.1 &  38.1 & 29.9 & 23.4 & 15.3 & 12.9 & 8.1\\
3$\times$10$^{-7}$  & 4 & 328 & 44.8 & 35.0 & 27.3 & 17.6 & 14.8 & 9.1 & 44.8 & 35.0 & 27.3 & 17.6 & 14.8 & 9.1  \\
10$^{-7}$ & 4 & 249 & 58.0 & 45.0 & 35.0 & 22.1 & 18.6 & 10.9 & 58.0 & 45.0 & 35.0 & 22.1 & 18.6 & 10.9 \\ 
\tableline
\tableline
\multicolumn{2}{c|}{M$_{p}$ from SB} & $T_{eff}$ (K)  & J  & H & K & L' & M & N \\
\tableline
\multicolumn{2}{c|}{1 M$_{J}$, ``cold start''} & 550 & 19.0 & 19.3 & 18.0 & 15.4 &13.0 & 12.0\\
\multicolumn{2}{c|}{1 M$_{J}$, ``hot start''}    & 820 &15.7 & 15.3 & 13.3 & 12.7 & 11.4 & 10.0 \\
\multicolumn{2}{c|}{10 M$_{J}$, ``cold start''}& 700 &17.3 & 16.8 & 17.2 & 14.1 & 12.6 & 11.9 \\
\multicolumn{2}{c|}{10 M$_{J}$, ``hot start''}& 2300 &9.5   & 8.7    & 8.1   & 7.8   & 8.3   & 7.5 \\
\tableline
\multicolumn{2}{c|}{GM Aur model} &  & J  & H & K & L' & M & N \\
\tableline
\multicolumn{2}{c|}{K5 V star with a disk} & & 3.2 & 2.7 & 2.5 & 2.1 & 1.9 & -0.1\\
\tableline

\end{tabular}
\end{center}
\end{table*}

 The absolute $J$, $H$, $K$, $L'$, $M$, and $N$-band magnitudes (magnitudes when the disk is at 10 pc away) of the SEDs are given in Table 1
 when the disk is observed face on. When the disk is not observed face on, the factor of $cos(i)$ needs to be multiplied to the luminosity to correct
 the absolute magnitudes for the geometric effect (Equation \ref{eq:ldisk}). 
The band passes are shown as the thin black curves at the top of each panel in Figure \ref{fig:SED}.
The total absolute magnitude of the planet-disk system can be calculated by
\begin{equation}
M=-2.5 \,{\rm log_{10}}\left(10^{-0.4 M_{p}}+10^{-0.4 M_{d}}\right)\,,
\end{equation}
where $M_{p}$ and $M_{d}$ are the absolute magnitudes of the planet and disk in Table 1 respectively. 
Absolute magnitudes for planets with other masses are given in Table 1 of SB.

\section{Magnetospheric Accretion}

\begin{figure*}[ht!]
\centering
\includegraphics[trim=0cm 0cm 0cm 0cm, width=0.45\textwidth]{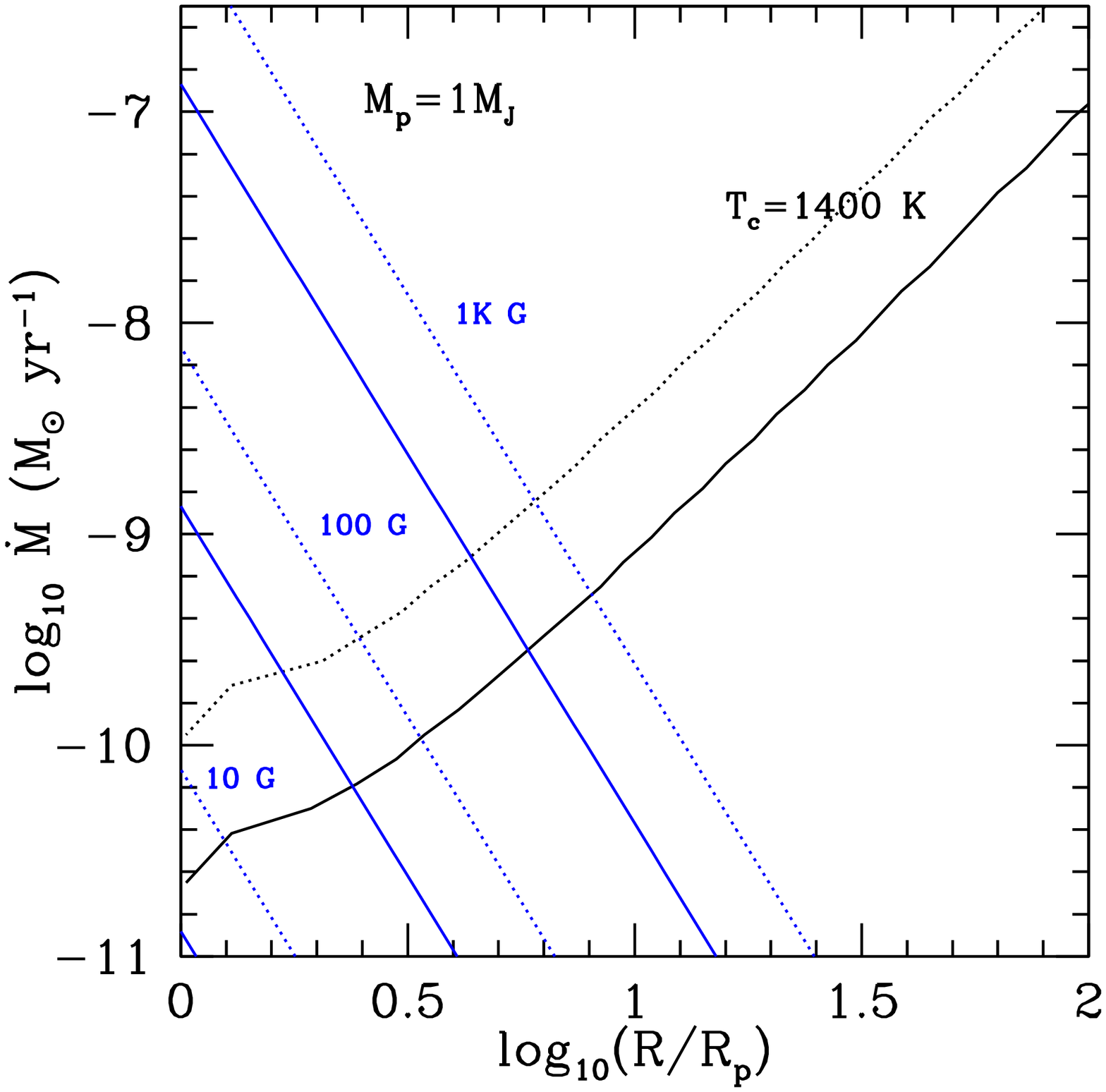}
\includegraphics[trim=0cm 0cm 0cm 0cm, width=0.45\textwidth]{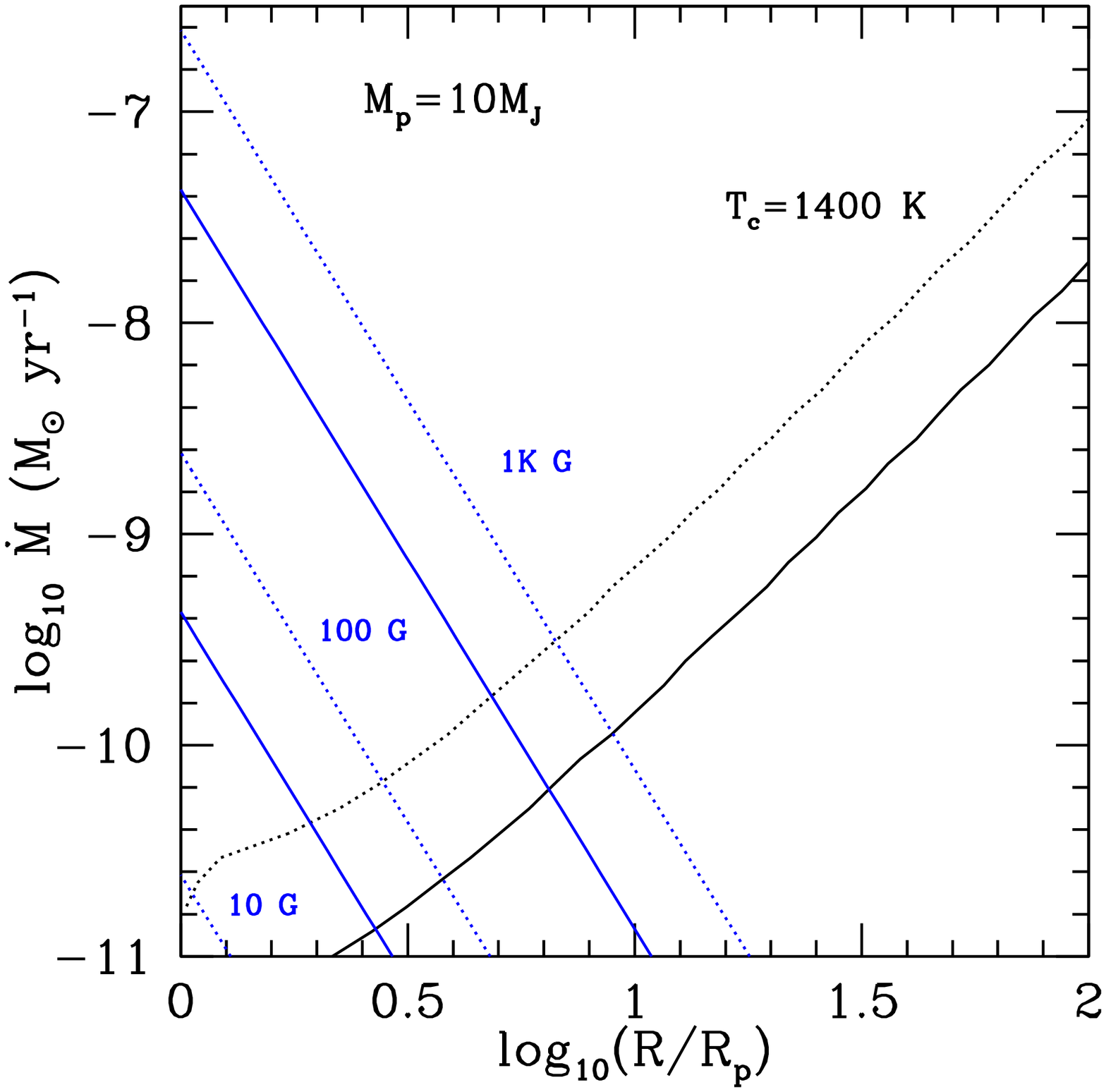} 
\caption{Blue lines: disk truncation radii due to planetary magnetic fields for disks accreting at
different accretion rates (Equation \ref{eq:RTRUN}). $M_{p}$ is1 M$_{J}$ in the left panel while $M_{p}$ is 10 M$_{J}$
in the right panel. 
The field strengths  at the equation of the planet surface are assumed to be
10, 100, and 1000 Gauss. Black curves: disk radii where the disk midplane temperature is 1400 K.
At smaller radii, the disk should be fully MRI active.
For both blue and black curves, the solid curves are derived by assuming $R_{p}=$R$_{J}$ while
the dotted curves assume  $R_{p}=$2 R$_{J}$.
}
\vspace{-0.1 cm} \label{fig:unstable1}
\end{figure*}

Accretion boundary layers and magnetospheric accretion are important components in the 
standard accretion disk picture. For a high accretion rate disk around a central object having a weak
magnetic field, the Keplerian rotating disk joins the slowly rotating central object through a boundary layer,
and  half of the accretion energy ($L_{acc}=GM_{p}\dot{M}/R_{p}$)
is released at the boundary  layer (Popham \& Narayan 1992). Assuming the boundary layer has a radial width
of disk scale heigh ($H\sim$0.1 $R_{p}$), the boundary layer has the typical effective temperature of
\begin{eqnarray}
T_{b,eff}&=&(\frac{L_{acc}}{\sigma 4\pi R_{p}H})^{1/4}\\
&=&7435 \,{\rm K}  \left(\frac{M_{p}}{{\rm M}_{J}}\right)^{1/4}\left(\frac{\dot{M}}{10^{-8}\msunyr}\right)^{1/4} \left(\frac{R_{p}}{\rm R_{J}}\right)^{-3/4}\,.
\end{eqnarray}
Thus the boundary layer would provide strong UV/optical flux.
However, we caution that, observationally, the simple boundary layer theory
overpredicts the X-ray emission from cataclysmic variables (Fertig et al. 2011) and UV emission from FU Orionis objects (Hartmann \etal 2011). 
For FU Orionis objects, Equation (\ref{eq:Fv}) (with slight modifications described after the equation), which has not included the boundary layer emission, still provides
the best match to the observed SED (Zhu \etal 2008, 2009; Hartmann \etal 2011). 

On the other hand, when the disk accretion rate is not so high and the planet magnetic 
fields are strong, the inner disk can be truncated 
by planet magnetic fields and the planet accretes through magnetospheric accretion (Lovelace \etal 2011).
Both dynamics and observational signatures of magnetospheric accretion have been relatively well studied for 
protostars (Romanova \etal
2008, Calvet \& Gullbring 1998, 
Muzerolle \etal 1998, 2001). { Dynamically, locking by the magnetosphere and the disk will spin down the rotation 
of the central objects. Unlike T Tauri stars that rotate very slowly compared to their breakup rates, Jupiter currently rotates
rapidly, implying Jupiter has little magnetosphere when it is surrounded by circumplanetary disks. } However, it is unknown
whether young exoplanets rotate very rapidly as Jupiter. Observationally,   
the SED signatures of magnetospheric accretion are different
from those of disk accretion. In this section, I will first calculate under what circumstances magnetospheric accretion can occur, then derive
the structure of the magnetosphere, and finally discuss the observational signatures of magnetospheric accretion.

The first-order estimate for the 
truncation radius is derived by equating the ram pressure
of a free-falling spherical envelope with the magnetic pressure of a dipolar field (Ghosh \& Lamb 1979),
\begin{equation}
\frac{R_{T}}{R_{p}}=\left(\frac{B_{p}^4R_{p}^5}{2GM_{p}\dot{M}^{2}}\right)^{1/7}\,,
\end{equation}
where $B_{p}$ is the magnetic field strength at the equator of the planet surface.
If the results are normalized  with the fiducial values of 
$B_{p}=100$G, $R_{p}=R_{J}$, $M_{p}=M_{J}$, and $\dot{M}=10^{-9}\msunyr$
we have
\begin{equation}
\frac{R_{T}}{R_{p}}=1.09\left(\frac{B_{p}}{100 G}\right)^{4/7}\left(\frac{R_{p}}{R_{J}}\right)^{5/7}\left(\frac{M_{p}}{M_{J}}\right)^{-1/7}\left(\frac{\dot{M}}{\dot{M_{-9}}}\right)^{-2/7}\,,\label{eq:RTRUN}
\end{equation}
where $\dot{M_{-9}}$ represents $10^{-9}\msunyr$. 
For fast rotators, the truncation radius can be 2 times the $R_{T}$ estimated above (Lovelace \etal 2011).

In Figure \ref{fig:unstable1}, blue curves are $R_{T}/R_{p}$ for 1 M$_{J}$
and 10 M$_{J}$ planet based on Equation (\ref{eq:RTRUN}). The solid curves are derived by assuming $R_{p}=$R$_{J}$,
while the dotted curves assume $R_{p}=$2$R_{J}$.
The magnetic field strength of young planets is unknown. Jupiter has an equatorial 
field strength of 4.28 G. With such a weak magnetic field, even slow accretion with $\dot{M}\gtrsim 10^{-11}\msunyr$
can crush the magnetosphere to the planet. 
Young Jupiters may have stronger magnetic field up to $\sim$60 G (S{\'a}nchez-Lavega 2004).
Young protostars can have field strength up to several kG (Johns-Krull \etal 1999).
Thus, the truncation radii for planets with $B_{p}$ of 10, 100, and 1 kG have been calculated.
Figure \ref{fig:unstable1} shows that magnetospheric accretion only occurs
when $\dot{M}\lesssim 10^{-9}\msunyr$ for 100 G magnetic field, consistent with Lovelace \etal (2011).

The structure of the magnetosphere around young planets can be calculated by following 
Hartmann \etal (1994), Calvet \& Gullbring (1998) and Muzerolle \etal (2003) which successfully
apply magnetospheric accretion theory
to  classical T Tauri stars (CTTS).
During magnetospheric infall, material reaches the planet surface at the free fall velocity of (Calvet \& Gullbring 1998)
\begin{eqnarray}
v_{s}&=&\left(\frac{2GM_{p}}{R_{p}}\right)^{1/2}\left(1-\frac{R_{p}}{R_{T}}\right)^{1/2}\\
&=&59 \,{\rm km}\, {{\rm s}}^{-1}\left(\frac{M_{p}}{{\rm M}_{J}}\right)^{1/2}\left(\frac{R_{p}}{{\rm R}_{J}}\right)^{-1/2}\zeta^{1/2}\,,\label{eq:vsplanet}
\end{eqnarray}
where
\begin{equation}
\zeta=1-\frac{R_{p}}{R_{T}}\,.
\end{equation}
This infall velocity is one order of magnitude smaller than the infall velocity onto a CTTS with solar mass and solar radius.

Knowing the velocity, the density of the magnetospheric accretion column can be estimated using the law of conservation of mass at the given accretion rate
\begin{equation}
\rho=\frac{\dot{M}}{f 4\pi R_{p}^{2}v_{s}}\,,
\end{equation}
so that
\begin{eqnarray}
n_{H}&=&10^{15}\,{\rm cm^{-3}}\left(\frac{\dot{M}}{\dot{M_{-9}}}\right)\left(\frac{M_{p}}{{\rm M}_{J}}\right)^{-1/2}\left(\frac{R_{p}}{{\rm R}_{J}}\right)^{-3/2}\nonumber\\
&&\zeta^{-1/2}\left(\frac{f}{0.01}\right)^{-1}\,,\label{eq:rhoplanet}
\end{eqnarray}
where $f$ is the filling factor of the accretion column on the planet surface and it ranges from
0.001 to 0.1 in CTTS (Calvet \& Gullbring 1998). This typical density in Equation (\ref{eq:rhoplanet}) 
is two orders of magnitude higher than
the typical density of the magnetosphere around a CTTS ($n_{H}\sim 10^{13}$ cm$^{-3}$) with $\dot{M}=10^{-8}\msunyr$ and $f=0.01$, which suggests 
that the emission from the magnetosphere of planets  will be more optically thick 
than that from the magnetosphere of CTTS. 

The supersonic magnetospheric infall leads to the shock formation at the planet surface.  In classical T Tauri stars (CTTS), the emission from the shock region is responsible
for  the ultraviolet excess and line veiling (Calvet \& Gullbring 1998). 
The total luminosity from the accretion shock is
\begin{equation}
L_{shock}=\zeta \frac{GM_{p}\dot{M}}{R_{p}}\,.\label{eq:lshock}
\end{equation}
When $R_{T}$ is much larger than $R_{p}$ or $\zeta\sim$1, almost entire accretion luminosity ($GM_{p}\dot{M}/R_{p}$) is released at the shock.
Since the accretion shock only has a small filling factor ($f$) on the stellar surface, the flux from the shock ($\mathscr{F}$) is
\begin{eqnarray}
\mathscr{F}&=&\frac{L_{shock}}{4\pi R_{p}^{2}f}=\frac{\zeta GM_{p}\dot{M}}{4\pi R_{p}^{3}f}\\
&=&1.73\times 10^{11} {\rm erg}\,{\rm cm}^{-2}\,{\rm s}^{-1}\left(\frac{M_{p}}{ {\rm M_{J}}}\right)\left(\frac{R_{p}}{{\rm R_{J}}}\right)^{-3}\nonumber\\
&&\left(\frac{\dot{M}}{\dot{M_{-9}}}\right)\zeta\left(\frac{f}{0.01}\right)^{-1}\,.\label{eq:fshock}
\end{eqnarray}
This high flux from the shock can heat up the planet photosphere below the shock. The ``heated photosphere''
has a temperature of $T_{hp}\sim (3 \mathscr{F}/4\sigma)^{1/4}$ (Calvet \& Gullbring 1998). With Equation (\ref{eq:fshock}), we have 
\begin{eqnarray}
T_{hp}&=&\left(\frac{3\zeta GM_{p}\dot{M}}{16 \sigma \pi R_{p}^{3}f} \right)^{1/4}\\
&=&6915\,{\rm K} \left(\frac{M_{p}}{ {\rm M_{J}}}\right)^{\frac{1}{4}}\left(\frac{R_{p}}{{\rm R_{J}}}\right)^{-\frac{3}{4}}
\left(\frac{\dot{M}}{\dot{M_{-9}}}\right)^{\frac{1}{4}}\zeta^{\frac{1}{4}}\left(\frac{f}{0.01}\right)^{-\frac{1}{4}}\label{eq:thpplanet}
\end{eqnarray}
This temperature is close to the typical value of $T_{hp}$ in CTTS.
When the circumplanetary disk has a low accretion rate of $10^{-10}\msunyr$, T$_{hp}$ is $\sim$ 3000 K which
is consistent with that derived by Lovelace \etal (2011).

\begin{figure*}[ht!]
\centering
\includegraphics[trim=0cm 0cm 0cm 0cm, width=0.9\textwidth]{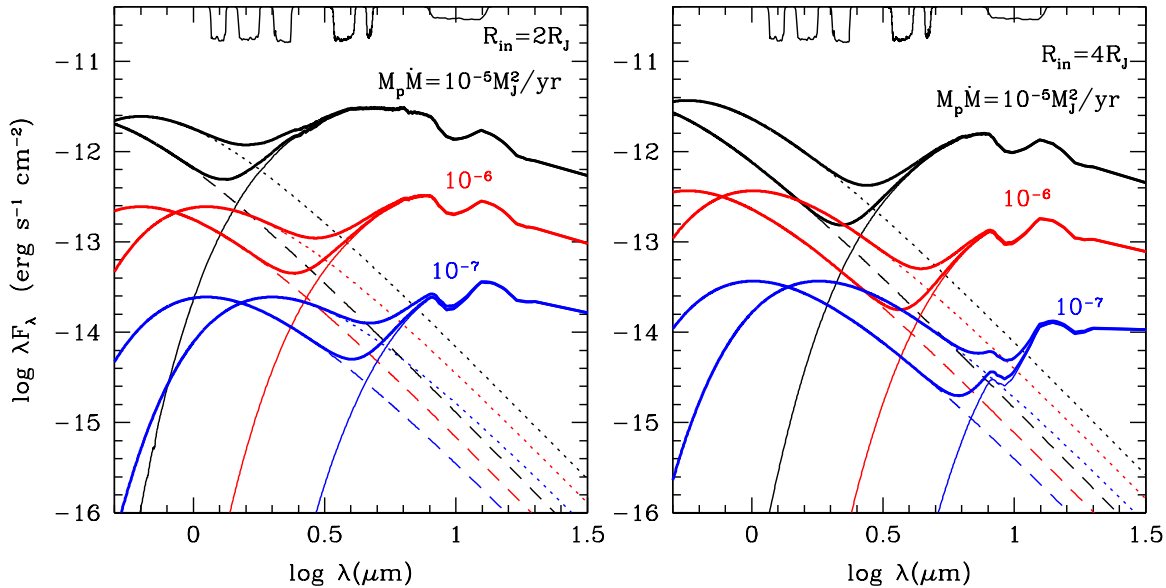} 
\vspace{-8.1 cm}
\caption{ Similar to Figure \ref{fig:SED} but including the emission from the heated photosphere due to the magnetospheric accretion. The disk is truncated
at $R_{in}$ and 
accretes to the planet (with a radius of R$_{J}$) through magnetospheric accretion. 
The thin solid, dotted, and dashed curves are the emission from the disk (the same as those shown in Figure \ref{fig:SED}), 
the heated photosphere with $f=0.1$, and the heated photosphere with $f=0.01$. The thick solid curves are the total emission adding both
disk emission and heated photosphere emission. 
Only disks with low accretion rates and large truncation radii are shown. }
 \label{fig:SEDshock}
\end{figure*}

With $T_{hp}$ given by Equation (\ref{eq:thpplanet}) and the assumed filling factor ($f$), we can calculate the SEDs of the heated photosphere
due to magnetospheric accretion by making the simplest assumption that the emission from the heated photosphere is a blackbody spectrum.
The SEDs of the heated photosphere, together with the SEDs of the accretion disk, are shown in Figure \ref{fig:SEDshock}. The truncation radius $R_{T}$ is also 
the inner radius of the accretion disk $R_{in}$. The planet radius is assumed to be R$_{J}$.
Two filling factors ($f$=0.1 and 0.01) have been assumed. Thus,    
the total SED from the planet-disk system has three components in the scenario of magnetospheric accretion: 
emission from the planet (red and blue curves in Figure \ref{fig:SED}), emission from the heated photosphere with a filling factor of $f$ on the planet surface
(dotted and dashed curves in Figure \ref{fig:SEDshock}), and emission from the accretion disk (thin solid curves in Figure \ref{fig:SEDshock}). 
Figure \ref{fig:SEDshock} shows that the hot heated photosphere can 
lead to strong optical and near-IR flux, and its SED can be even hotter and brighter than the SED from the planet.
The blackbody assumption for the heated photosphere is highly simplified. The SED from the heated photosphere 
would be closer to that from an atmosphere of the given $T_{hp}$. Since the total SED consists of these three components,
the lines emitted by each component
could be shallower/veiled due to the emission from other components.

Besides the UV/optical excess and line veiling, another observational signature of
magnetospheric accretion is emission of atomic lines  (e.g. Balmer, Paschen and Na D lines).
Estimating the line flux is more difficult since these lines come from the whole magnetosphere, and
we know little on the heating mechanisms there (e.g. heating by shocks, MHD waves). If we adopt the typical temperature
values of the magnetosphere constrained from CTTS ($\sim$ 8000 K), we can roughly estimate the line flux of H$_{\alpha}$. 
At the typical scale of R$_{\odot}$ across the magnetosphere of CTTS,
H$_{\alpha}$
is optically thick with 8000 K and $n_{H}>10^{12}$ cm$^{-3}$ (Storey \& Hummer 1995). Although the magnetosphere of planets
is smaller, the density is significantly higher (Equation \ref{eq:rhoplanet}) so that H$_{\alpha}$ is still optically thick. At $n_{H}\gtrsim 10^{12}-10^{13}$  cm$^{-3}$,
the lower levels of H are controlled by collisions (Hartmann \etal 1994), so that the source function for H$_{\alpha}$ is the Planck function.
Thus, considering
the size of the magnetosphere is $4\pi R_{T}^2$, the line 
luminosity would be
\begin{eqnarray}
L_{H_{\alpha}}&=&\pi B_{\nu}(T=8000 K)4\pi R_{T}^2 \times \frac{v_{s}}{c}\nu_{H_{\alpha}}\\
&=& 4.7\times 10^{-6} {\rm L_{\odot}}\left(\frac{R_{T}}{{\rm R_{J}}}\right)^{2}\left(\frac{v_{s}}{59 \,{\rm km}\,{\rm s^{-1}}}\right)\label{eq:Lhplanet}\,,
\end{eqnarray}
where I assume that the line is a box of width $v_{s}\nu_{H_{\alpha}}/c$ (broadened by the infall) and the height of $\pi B_{\nu_{H_{\alpha}}}(T=8000 K)$.
Since $\pi B_{\nu_{H_{\alpha}}}$ is the flux at the line center,  Equation (\ref{eq:Lhplanet}) is the upper limit of the line flux. 
Another reason for this estimate to be the upper limit is that, at high densities ($n_{H}\sim 10^{15}$  cm$^{-3}$), H can be in local thermal equilibrium and
the continuum emission can be as strong as the line emission. 
Nevertheless,
the typical value in Equation (\ref{eq:Lhplanet}) is three orders of magnitude lower than the typical value from CTTS ($5\times 10^{-3}$L$_{\odot}$ 
in Herczeg \& Hillenbrand 2008; Rigliaco \etal 2012) due to one order of magnitude smaller
truncation radius and one order of magnitude smaller infall velocity. 

{ Our estimate of $L_{H_{\alpha}}$ upper limit (Equation \ref{eq:Lhplanet}) is consistent with measurements of accreting late M type brown dwarfs by Zhou \etal (2014). 
The smallest object in Zhou \etal (2014) is GSC 06214-210b with a radius of 1.5 $R_{\rm J}$ and $L_{H_{\alpha}}\sim8\times10^{-6}$ 
L$_{\odot}$. The other two objects in Zhou \etal (2014) (GQ Lup b and DH Tau b) have radii of $5$ and $2.7$ R$_{\rm J}$
with $L_{H_{\alpha}}$= $\sim2\times10^{-6}$ and $7.5\times 10^{-7}$  L$_{\odot}$. These $L_{H_{\alpha}}$ values
are all close or smaller than our upper limit given by Equation (\ref{eq:Lhplanet}).}

Equation (\ref{eq:Lhplanet})  does not show the known correlation between 
$\dot{M}$ and $L_{H_{\alpha}}$ (Herczeg \& Hillenbrand 2008; Rigliaco \etal 2012). 
We suggest then that the filling factor, or total azimuthal coverage, of the accretion flow provides the correlation between the accretion 
rate and the emission-line flux. Realistically, the flow is not azimuthally axisymmetric but is probably separated into discrete magnetic 
flux tubes or ``funnel flows''. This is especially likely if the magnetic field is tilted with respect to the stellar rotation axis, such that 
accretion will occur preferentially where the field lines bow inward toward the star. In this scenario, higher accretion rates result 
in a larger number of (or wider) discrete flows, providing a larger emitting region and hence stronger line fluxes. There must be 
a relatively symmetric distribution of flows around the star, so that the full range of red- and blueshifted velocities are in emission 
(for instance, a single discrete flow aligned along the line of sight to the star would result in a profile with a highly redshifted peak and little emission blueward of line center, which has not been observed in CTTS, Muzerolle \etal 2001).
Overall Equation (\ref{eq:Lhplanet}) 
suggests that H$_{\alpha}$ luminosity is proportional to the square of the magnetosphere radius so that H$_{\alpha}$
luminosity may decrease significantly when the objects are in the planet mass regime due to their weak magnetic fields.

\section{Discussion}
\subsection{Caveats}
{ We have made a major simplification by assuming that the circumplanetary disk accretes at  a constant rate throughout
the whole disk. On the other hand, recent 3-D simulations (Machida  \etal 2008, Tanigawa \etal 2012, Szulagyi \etal 2014) 
have suggested that gas enters the circumplanetary disk almost vertically from the circumstellar disk. This vertical infall
can even be significant at two Jupiter radii (Szulagyi \etal 2014). Martin \& Lubow (2011) has suggested that the standard accretion
disk picture only stands inward of where the vertical entrainment occurs. The spectrum of radiation from the region between the mass infall
radius and the outer edge of the disk corresponds to that of a decretion disk (Equation 66 in Martin \& Lubow 2011) with
$T_{eff}^4\propto R^{-3.5}$ instead of $T_{eff}^4\propto R^{-3}$ for an accretion disk (Pringle 1981, Lee 1991). Furthermore,
the mass infall may not occur at a single radius, but instead a wide range of radii. Thus, assuming the disk accretion 
rate at the disk 
radius $R$ equals to the integrated infall rate beyond $R$, the disk accretion rate decreases with $R$. Both of the decretion
disk structure and decreasing $\dot{M}$ lead to steeper temperature profile with radii. To explore this effect, we have
calculated the SED of a disk  with $T_{eff}^4\propto R^{-4}$ by multiplying Equation \ref{eq:Fv} with $1.36 R_{i}/R$ (again, 
the temperature at $R<1.36 R_{i}$ is assumed to be the same as that at $R=1.36 R_{i}$). Assuming $R_{i}=R_{J}$ and
$M\dot{M}=10^{-5}$M$_{\rm J}^2$/yr, the obtained J, H, K, L', M, N magnitudes are 9.9, 9.8, 9.0, 7.6, 7.1, 6.0, instead of
9.3, 9.2, 8.3, 6.6, 6.0, 4.5 in Table 1. Thus, the disk becomes slightly fainter and the 
slope of the SED becomes slightly steeper due to the steeper temperature profile.}   

The standard accretion disk model  is based on the assumption that local turbulence is responsible for both 
angular momentum transport and energy dissipation. However, circumplanetary disks feel strong
tidal forces from the central star. This tidal force can drive global spiral arms and shocks in circumplanetary disks,
leading to accretion. 2-D and 3-D simulations (Rivier \etal 2012, Szulagyi \etal 2014) have shown that
the accretion induced by tidal forces is $\lesssim 2\times10^{-10} \msunyr$ assuming that dissipation in the circumplanetary
disk leads to a steady accretion. However, shock dissipation is unlikely to be uniform throughout the circumplanetary disk
and mass may pile up in the disk by shock dissipation. In this case, local turbulence is still needed for accretion onto the planet. 

Another mechanism for global angular momentum transport is the disk wind. The wind can lead to significantly less energy dissipation
compared with the dissipation in viscous models.
Recent MHD simulations by Gressel \etal (2013)
have suggested that a magnetocentrifugal wind develops in circumplanetary disks. Future work is required to know the
fraction of angular momentum and energy carried by disk wind versus the local turbulence. 

Regarding the SED calculations, when
 $\dot{M}\sim 10^{-9}-10^{-10}\msunyr$ with $\alpha=0.01$, our assumption that the disk is optically thick marginally stands 
within the dust sublimation radius where the opacity is dominated by molecular opacity. 
In this case, the disk may have a different temperature structure than that obtained through the
grey atmosphere assumption.  Cooling through molecular lines may be quite important in this case. 

Finally, we discuss how mass can be transported in circumplanetary disks.
The disk needs to be hot enough to sustain MRI.
When the disk accretion rate is too low, viscous heating may not provide enough thermal ionization to sustain MRI. 
By assuming $\alpha=0.01$,  the disk thermal structure has been calculated (following Zhu \etal 2009) 
and we plot the radius where the disk midplane
temperature reaches 1400 K in Figure \ref{fig:unstable1}. As shown, when $\dot{M}\lesssim 10^{-10}-10^{-11}\msunyr$, 
MRI cannot be sustained by thermal ionization in the disk. The disk may need to 
rely on non-thermal ionization at the disk surface to sustain MRI, which is quite uncertain (Fujii \etal 2011, 2014, Turner \etal 2014). 

\subsection{Observational Signatures}

The disk SED calculations suggest that, at near-IR wavelengths ($J$, $H$, $K$ bands), a 
moderately accreting circumplanetary disk ($\dot{M}\sim 10^{-8}\msunyr$) 
can be as bright as a late M-type brown dwarf or a 10$M_{J}$ planet with a ``hot start'', since a late M-type dwarf  (Basri \etal 2000)
and a 10 M$_{J}$ planet with a ``hot start'' (SB) both have T$_{eff}\sim$ 2300 K and the radius of Jupiter. 
This result implies that recently discovered high mass young planet/brown dwarf candidates that are in protoplanetary disks (Neuh{\"a}user et 
al. 2005; Itoh \etal 2005;  Kraus \& Ireland 2012; Quanz \etal 2013) could also 
be accreting circumplanetary disks around low mass planets.

To distinguish  the accretion disk around a low mass planet (e.g., 1 M$_{J}$) 
from a brown dwarf or a hot high mass planet, it is crucial   to obtain the photometry at mid-IR bands ($L'$, $M$, $N$ bands)  for these objects.
Since the total flux of an accretion disk is from different disk annuli having a wide range of temperatures,
the SEDs of accretion disks fall off slower towards longer wavelengths than the Rayleigh- Jeans tail of the SEDs of brown dwarfs or planets  (Figure \ref{fig:SED}).

Searching for circumplanetary disks at radio wavelengths is also appealing due to 
ALMA's high sensitively (Isella et al. 2014). However, dust in circumplanetary disks may be
depleted due to particle trapping at the planet opened gap edge (Rice \etal 2006, Zhu \etal 2012)
or depleted due to radiation pressure from the accreting planets (Owen 2014). 

If the planet can have a strong magnetic field, the field can truncate the  
 accreting circumplanetary disk to produce 
other accretion signatures, including
optical/UV excess from the heated photosphere due to accretion shocks, line veiling, and $H_{\alpha}$ 
emission from magnetospheric accretion. Gyrosynchrotron radiation modulated at a short period
could also be a good indicator for magnetospheric accretion around planets (Lovelace \etal 2011). 
The heated photosphere produces a high temperature spectrum but from a small emitting area with the filling factor of $f$ on the planet surface,
which is distinguishable from the emission from the planet and the disk.
Close \etal (2014) have detected strong  $H_{\alpha}$ emission within cavities of transitional disks,
and Zhou \etal (2014) has suggested that $H_{\alpha}$ flux can be as high as $50\%$
of the accretion luminosity for disks around late M or L type brown dwarf. 
On the other hand, we caution that $H_{\alpha}$ line flux sensitively depends
on the size of the magnetosphere, which could be small if the magnetic field of the planet is weak (\S 4).

{ The outer region of the circumplanetary disk, which is close to the Roche sphere, may also experience strong
dissipation due to shock formation by the infall or Kelvin-Helmholtz instability at the position where infall meets the disk (similar
to the "hot spot" in cataclysmic variables).
The observational  signatures of the outer disk need further study. }

\subsection{Application to HD 169142}
During the preparation of this manuscript, Reggiani \etal (2014) and Biller \etal (2014) reported the discovery of a companion
around the transitional disk star  HD 169142. Reggiani \etal (2014) found that the companion has L'=12.2 $\pm$ 0.5 and J$>$13.8.
Assuming the distance of 145 pc, the absolute magnitude is L'=6.4 $\pm$ 0.5 and J$>$8.0. Assuming the circumplanetary disk
has an inner radius of R$_{J}$ and is viewed face on, our Table 1 suggests that $M_{p}\dot{M}\sim 10^{-5}$M$_{J}^{2}$/yr with the prediction of  apparent
magnitudes: M=11.6$\pm$ 0.5 and N=10.1$\pm$ 0.5. Assuming the inner radius of 4 R$_{J}$, Table 1 suggests  
$M_{p}\dot{M}\sim 5\times 10^{-5}$M$_{J}^{2}$/yr with the prediction of  apparent
magnitudes: M=11.3$\pm$ 0.5 and N=9.5$\pm$ 0.5.
 If $M_{p}=1-10$ M$_{J}$, the circumplanetary disk would accrete at 10$^{-8}-10^{-9} \msunyr$, which is reasonable. 
 
Biller \etal (2014) report the non-detection of H$_{\alpha}$ in this system (Follette \etal In prep).
Using our Figure \ref{fig:unstable1}, the non-detection could imply that the magnetic field 
around the planet is weak (e.g. B$<$100 G) and the disk accretion rate is high (e.g., $\dot{M}>10^{-10}-10^{-9}\msunyr$), which is
consistent with the accretion rates obtained from L'.  
 
\section{Conclusion}
We calculate the spectral energy distributions (SEDs) of  circumplanetary disks  accreting at a variety of accretion rates. 
We find that the circumplanetary disks that are even accreting at  $10^{-10}\msunyr$ around a 1 M$_{J}$ planet can be brighter than the planet itself.
At near-IR wavelengths ($J$, $H$, $K$ bands), a 
moderately accreting circumplanetary disk ($\dot{M}\sim 10^{-8}\msunyr$) has a maximum temperature of $\sim$2000 K and
is as bright as a late M-type brown dwarf or a 10 M$_{J}$ planet with a ``hot start''. 
To distinguish  the accretion disks around low mass planets (e.g., 1 M$_{J}$) 
from brown dwarfs or hot high mass planets, it is crucial  to obtain the photometry at mid-IR bands ($L'$, $M$, $N$ bands)  since 
the disks' SEDs have flatter slopes towards longer wavelengths than the Rayleigh-Jeans tail of the SEDs of brown dwarfs or planets. 
If  young planets have strong magnetic fields ($\gtrsim$100 G), the magnetic fields may truncate the slowly accreting circumplanetary disks ($\dot{M}\lesssim$10$^{-9}\msunyr$) 
and lead to magnetospheric accretion,
which can provide additional accretion signatures (e.g. UV/optical excess, line veiling and line emission).
James Webb Space Telescope will provide enough sensitivity and wavelength coverage (near to mid-IR) to 
study circumplanetary disks in future.

\acknowledgments
Z.Z. acknowledges support by
NASA through Hubble Fellowship grant HST-HF-51333.01-A
awarded by the Space Telescope Science Institute, which is
operated by the Association of Universities for Research in Astronomy, Inc., for NASA, under contract NAS 5-26555.
Z.Z. thanks Lee Hartmann for helpful discussions and encouragement to complete this work, and Nuria Calvet for numerous discussions
on the basics of magnetospheric accretion. Z.Z. thanks the referee Steve Lubow for insightful comments to improve the paper. 
Z.Z. also thanks  Adam Burrows, Kate Follette, James Muzerolle, Sascha Quanz, Roman Rafikov, Jim Stone, and Neal Turner 
 for helpful discussions, and David Spiegel for providing
transmission functions for different bands.

\end{document}